\documentclass[showpacs,preprint,11pt,nofootinbib,superscriptaddress,citeautoscript,eqsecnum]{revtex4}
 \usepackage{amsmath,amsfonts,amssymb}
\usepackage{color}
\definecolor{darkblue}{rgb}{0.0,0.0,0.8}
\definecolor{darkred}{rgb}{1,0.5,0.7}
\definecolor{Brown}{cmyk}{0, 0.8, 1, 0.6}
\usepackage{setspace}
\usepackage{hyperref}
\hypersetup{colorlinks=true, citecolor=darkblue, linkcolor=Brown,  urlcolor=darkblue}

\newcommand{\ba}{\begin{eqnarray}}
\newcommand{\ea}{\end{eqnarray}}
\newcommand\be{\begin{equation}}
\newcommand\ee{\end{equation}}

\input{epsf.sty}

\begin{document}

 \title{The stability and gravitational Newtonian limit of a modified Randall-Sundrum model}

\author{Shahrokh Parvizi}
\email{parvizi@modares.ac.ir}
\affiliation{Department of Physics, School of Sciences, Tarbiat Modares University, P.O. Box 14155-4838, Tehran, Iran}

\author{Mojtaba Shahbazi}
\email{Mojtaba.Shahbazi@modares.ac.ir}
\affiliation{Department of Physics, School of Sciences, Tarbiat Modares University, P.O. Box 14155-4838, Tehran, Iran}

 \date{\today}

\begin{abstract}
For a modified Randall-Sundrum model [Phys. Rev. D 88 (2013) 025048], the graviton equations are derived and the mass spectrum found. The latter includes a massless graviton and a continuum mass with a gap. There is no negative mass-squared in the spectrum, so the model is stable. The gravitational Newtonian limit is obtained with an exponentially suppressed modification from extra dimension.
\end{abstract}


\maketitle
\section{Introduction}
Extra dimensional brane-world models firstly introduced in \cite{ArkaniHamed:1998rs} and then in \cite{Randall:1999ee} to resolve the hierarchy problem in fundamental interactions including gauge fields and gravity. The basic idea of these models and their followers is an assumption that matter and  gauge  fields are confined on a 3-dimensional brane embedded in a higher dimensional spacetime while the gravity, by definition, can travel in all dimensions. This is compatible with observations provided either the extra dimension volume to be in order of TeV scale as suggested in 
\cite{ArkaniHamed:1998rs} or somehow warped over the brane as in \cite{Randall:1999ee}. 

The mentioned assumption is supported by string theory in which (D-)branes  are defined to be where open strings ended. The latter correspond to the standard model fields in the low energy limit. In contrast, closed strings can propagate into extra dimension(s) off the brane and in the low energy limit correspond to gravitons. Despite of this justification, it is important to understand the mechanism of field localization on the brane in the low energy scales and independent of the string theory.  This was not very satisfactory and all Standard Model's particles could not be trapped into the brane in the original 5-dimensional version of this model, though there are  successes by some 6-dimensional models \cite{all}\cite{gauge}\cite{six}.

 In this regards, in \cite{prd}, a 5-dimensional  modified Randall-Sundrum (MRS) model was proposed and shown that it improves the field localization behavior on the brane. The model looks like very similar to the original RS and even two geometries are locally the same, but it was shown that they differ globally \cite{prd}. An important difference is that, in contrast to the RS model, the new one does not contain any 5-dimensional cosmological constant.
 
 Regarding progress in the field localization by proposing \cite{prd}, it is worth examining the effective 4-dimensional   gravitational behavior of this model. In this article, we investigate the gravitational perturbation of the background and find the graviton spectrum. We show that it excludes negative mass modes which signing stability of the model. Moreover its Newtonian gravity limit is studied. 

The organization of this paper is as follows. In section 2 we introduce the metric and remind its differences with the RS. In section 3, we perturb the background and find the graviton mass spectrum. In section 4, we study the Newtonian limit and conclude in section 5.

\section{The background metric}
Consider the  5-dimensional Einstein-Hilbert gravity action  as in the following,
\begin{equation}
S_{g}=\frac{1}{2\kappa^2}\int d^5x\sqrt{|G|} R  + S_{brane}\label{action}
\end{equation}
in which $S_{{\text brane}}$ is a localized action of brane(s).
Based on this action, the following metric can be introduced as a solution to the equations of motion which containes a 3-brane at the origin of fifth dimension \cite{prd}:
\begin{equation}\label{r-metric}
ds^2=e^{-2k|r|} \eta_{AB}dx^A dx^B=e^{-2k|r|} (\eta_{\mu\nu}dx^\mu dx^\nu+dr^2)
\end{equation}
where $r$ is the extra dimension varying from $-\infty$ to $+\infty$ and the Latin letters are from 1 to 5 and Greek letters (brane coordinates) from 1 to 4. In addition; $\eta_{\mu\nu}$ is Minkowskian metric with  $\eta_{\mu\nu}=diag(-1,1,1,1)$. Henceforth, we use the nomination of \cite{prd} and call (\ref{r-metric}) the r-metric.

It is worth recalling the well-known Randall-Sundrum (RS) metric as \cite{RS2}
\begin{equation}\label{RS2}
ds_{RS}^2=e^{-2k|z|} \eta_{\mu\nu}dX^\mu dX^\nu+dz^2
\end{equation}
At the first sight it seems that the r-metric (\ref{r-metric}) can be converted to the RS metric (\ref{RS2}), by the following coordinate transformation \cite{prd},
\begin{eqnarray}
e^{-k|r|}&=&1-k|z|\nonumber\\
dx^\mu&=&\frac{e^{-k|z|}}{1-k|z|}dX^\mu
\end{eqnarray}
However, these two metric can not be the same spacetime for at least two reasons \cite{prd}: Firstly, the above coordinate transformation is singular at $z=1/k$, so it can not be extended to all space. Secondly, a global coordinate transformation need to be an exact differential as $dx^\mu=\partial x^\mu/\partial X^\nu dX^\nu + \partial x^\mu/\partial z dz$ which is manifestly not the case. Moreover, it is shown in \cite{prd} that the consistent Einstein equation including the energy-momentum of the brane requires the cosmological constant to be zero as we already set in (\ref{action}), in contrast to the Randall-Sundrum model which contains a negative 5-dimensional cosmological constant. Furthermore the field localization of different spin fields gets improvement compared to the original Randall-Sundrum model \cite{prd}. 


\section{Field equations and Graviton modes}
To derive Einstein's field equations for the matter of convenience, different coordinate system is adopted:
\begin{equation}
 ds^2=e^{2\sigma}(\eta_{\mu\nu}dx^\mu dx^\nu)+dy^2 \label{metric}
\end{equation}

with $ \sigma=\ln(1-k|y|)$. 
By inserting (\ref{metric}) in the Einstein's equation:
\begin{eqnarray}
\kappa^2T_{\mu\nu}&=& g_{\mu\nu}(3\ddot{\sigma}+6\dot{\sigma}^2)  \label{e1}\\
\kappa^2T_{55}&=& 6\dot{\sigma}^2\label{e2}
\end{eqnarray}
where dot represents derivative with respect to the $y$~extra dimension.

To find graviton modes in the MRS model, metric would be purturbed around the fixed background (\ref{metric}) as follows:
\begin{equation}
G_{AB}=g_{AB}+h_{AB} \label{pertur}
\end{equation}
%
Plug (\ref{pertur})~into (\ref{action}) and keep up to second order in $h$, then the variation due to $h^{AB}$, gives the linearized equation of motion \cite{RS2}:
\begin{eqnarray}
& &-\frac{1}{2}\nabla^T\nabla_Nh_{MT}-\frac{1}{2}\nabla^T\nabla_Th_{MN}+\frac{1}{2}\nabla_M \nabla_N h+\frac{1}{2}g_{MN}\nabla^T \nabla^S h_{TS}-\frac{1}{2}\nabla^T \nabla_T h \nonumber\\
& &+\kappa^2[T_{MT}h^T_N+T_{NT}h^T_M-\frac{1}{2}T_{MN}h-\frac{1}{2}g_{MN}T_{TS}h^{TS}-\frac{T^S_S}{6}(2h_{MN}-g_{MN}h)] \nonumber\\
& &-\frac{1}{2}\kappa^2\mathcal{L}~ h_{MN}=0 \label{em}
\end{eqnarray}
where  $h=h^A_A$ and $T_{MN}$ is from (\ref{e1}) and (\ref{e2}). It could be shown that adopting the unitary gauge is feasible \cite{effective}:
\begin{equation}
h_{\mu5}=0~~ \&~~  h_{55}=f(x^\mu)e^{n\sigma}:=\varphi
\end{equation}
where $n$ is a constant. By rewriting the equation of motion (\ref{em}) for different components we get:

\begin{align}
\mu\nu-component:~&\frac{1}{2}(\partial_\rho{\partial^\rho{h_{\mu\nu}}}-\partial_\mu{\partial^\rho{h_{\rho\nu}}}-\partial_\nu{\partial^\rho{h_{\rho\mu}}}+\ddot{h}_{\mu\nu})+4\dot{\sigma}^2h_{\mu\nu}+\frac{1}{2}\partial_\mu{\partial_\nu{\tilde{h}}}+\frac{1}{2}\partial_\mu{\partial_\nu{\varphi}} \nonumber\\
&+\frac{1}{2}g_{\mu\nu}[\partial^\rho{\partial^\sigma{h_{\rho\sigma}}}-\partial_\rho{\partial^\rho{\tilde{h}}}-\ddot{\tilde{h}}-4\dot{\sigma}\dot{\tilde{h}}-\partial_\mu{\partial_\nu{\varphi}}-6\dot{\sigma}^2\tilde{h}+3\dot{\sigma}^2\varphi(2+n)] \nonumber\\
&+\ddot{\sigma}(\frac{3}{2}g_{\mu\nu}\varphi+2h_{\mu\nu}-\frac{3}{2}g_{\mu\nu}\tilde{h})-(3\ddot{\sigma}+6\dot{\sigma}^2)h_{\mu\nu}=0 \label{mn}\\
\mu5-component:~&\frac{1}{2} \partial_5{(\partial_\mu{\tilde{h}}-\partial^\nu{h_{\mu\nu}})}-\frac{3}{2}\dot{\sigma}\partial_\mu{\varphi}=0 \label{m5}\\
55-component:~&\frac{1}{2}(\partial^\mu{\partial^\nu{h_{\mu\nu}}}-\partial_\mu{\partial^\mu{\tilde{h}}})-\frac{3}{2}\dot{\sigma}\dot{\tilde{h}}+(13+2n)\dot{\sigma}^2\varphi-3\dot{\sigma}^2\tilde{h}=0\label{55},
\end{align}
in which $\tilde{h}=g^{\mu\nu}h_{\mu\nu}$. Eq.s (\ref{mn}),(\ref{m5}) and (\ref{55}) are a system of coupled partial differential equations which can be solved if one decouples them. To do this, we use the following tensor decomposition \cite{wein}:
\begin{equation}
h_{\mu\nu}=Eg_{\mu\nu}+B_{,\mu\nu}+C_{\mu,\nu}+C_{\nu,\mu}+D_{\mu\nu}\label{de}
\end{equation}
where $E$~and~$B$~are scalars (called radion in the context of RS model), $C_{\mu}$~ is a divergenceless vector and $D_{\mu\nu}$~ a divergenceless-traceless tensor (graviton). Then inserting in equations of motion one finds,
\begin{align}
\mu\nu-component:~&\frac{1}{2}\Big(-2E_{,\rho}^{\rho}g_{\mu\nu}-C_{\mu,\nu\rho}^{~~~\rho}-C_{\nu,\mu\rho}^{~~~\rho}+D_{\mu\nu,\rho}^{~~~\rho}+2E_{,\mu\nu}+(Eg_{\mu\nu})_{,55}+B_{,\mu\nu55}\nonumber \\
&+C_{\mu,\nu55}+C_{\nu,\mu55}+D_{\mu\nu,55}\Big)+\frac{1}{2}\varphi_{,\mu\nu}+4\dot{\sigma}^2(-2Eg_{\mu\nu}+B_{,\mu\nu}+C_{\mu,\nu}+C_{\nu,\mu} \nonumber\\
&+D_{\mu\nu})+\frac{1}{2}g_{\mu\nu}\Big(-4\ddot{E}-B_{,\rho55}^{\rho}-4\dot{\sigma}(4\dot{E}+B_{,\rho5}^{\rho})-\varphi_{,\rho}^{~\rho}-6B_{,\rho}^{~\rho}\dot{\sigma}^2 \nonumber \\
&+3\dot{\sigma}^2\varphi (2+n)\Big)+\ddot{\sigma}\Big(\frac{3}{2}g_{\mu\nu}\varphi-4g_{\mu\nu}E+2B_{,\mu\nu}+2C_{\mu,\nu}+2C_{\nu,\mu}+2D_{\mu\nu}\nonumber\\
&-\frac{3}{2}g_{\mu\nu}B_{,\rho}^{~\rho}\Big)=0\\
\mu5-component:~&\frac{3}{2}\dot{E}_{,\mu}-\frac{3}{2}\dot{\sigma} e^{n\sigma}f_{,\mu}-\frac{1}{2}\dot{C}_{\mu,\nu}^{~~\nu}=0 \\
55-component:~&-\frac{3}{2}\dot{\sigma}g^{\mu\nu}\dot{B}_{,\mu\nu}+(13+2n)\dot{\sigma}^2 e^{n\sigma}f-\frac{3}{2}E_{,\mu}^{~\mu}-12\dot{\sigma}^2E-6\dot{\sigma}\dot{E}=0
\end{align}

One of the advantages of the decomposition (\ref{de}) is making a distinction between the degrees of freedom of the model.
Since we are interested in studying the effects of graviton,  we can consistently set equal zero all parts of the decomposition (\ref{de}) but the tensorial part $D_{\mu\nu}$. Having done that, the only non-trivial equation would be the $\mu\nu-component$~equation of motion:
\begin{equation}
\partial_\rho{\partial^\rho{D_{\mu\nu}}}+\ddot{D}_{\mu\nu}-(4\dot{\sigma}^2+2\ddot{\sigma})D_{\mu\nu}=0 \label{g}
\end{equation}

Let us solve (\ref{g}) by two boundary conditions. The first one is, 
\begin{equation}
\lim_{y\rightarrow\frac{1}{k}}D_{\mu\nu}(x,y)=0
\end{equation}
and the second one is canceling out coefficients of delta Dirac in (\ref{g}).

By imposing the Fourier transform along the brane coordinates on eq. (\ref{g}):
\begin{equation}
+p^2\tilde{D}_{\mu\nu}-\ddot{\tilde{D}}_{\mu\nu}+(4\dot{\sigma}^2+2\ddot{\sigma})\tilde{D}_{\mu\nu}=0 \label{g1}
\end{equation}
where $p^2=g^{\mu\nu}p_\mu p_\nu=-m^2$. (\ref{g1}) is called Schrodinger-like equation and can be solved to find,
\begin{equation}
\tilde{D}_{\mu\nu}=w_{\mu\nu}e^{n_1\sigma}+u_{\mu\nu}e^{n_2\sigma} \label{a}
\end{equation}
where $w_{\mu\nu}$ and $u_{\mu\nu}$ are some constant tensors and:
\begin{equation}
n_1=\frac{1+\sqrt{(\frac{2\hat{p}}{k})^2+9}}{2}\; , \qquad
n_2=\frac{1-\sqrt{(\frac{2\hat{p}}{k})^2+9}}{2}
\end{equation}
with $\hat{p}^2:=\eta^{\mu\nu}p_\mu p_\nu$. Considering the second boundary condition results in:
\begin{align}
&\Big( w_{\mu\nu}(2-n_1)+u_{\mu\nu}(2-n_2)\Big) \delta(y)=0 \nonumber\\
&w_{\mu\nu}=-\frac{3+a}{3-a}u_{\mu\nu}:=Fu_{\mu\nu} \label{c}
\end{align}
where
\begin{equation}
F=-\frac{3+a}{3-a}\; , \qquad
a=\sqrt{(\frac{2\hat{p}}{k})^2+9}
\end{equation}
The first boundary condition in which the solution (\ref{a}) must vanish at $y=\frac{1}{k}$~ makes four divisions:
\begin{enumerate}
\item $m^2=0$ then $n_1=2$, $n_2=-1$ and by condition (\ref{c}), $u_{\mu\nu}=0$. So $w_{\mu\nu}$ can be found by normalizing the solution as follows, 
\begin{align*}
\int_{-1/k}^{1/k}e^{-\sigma}{}& D_{\mu\nu}^2dy = 1 \nonumber\\
|w_{\mu\nu}|^2\int e^{(2n_1-1)\sigma}dy {}&= |w_{\mu\nu}|^2\int(1-k|y|)^3dy=1\nonumber\\
|w_{\mu\nu}|^2 {}&= 2k
\end{align*}
For later use we find,
\begin{equation}
D_{\mu\nu}^2\Big|_{y=0}=2k \label{1}
\end{equation}
\item Either $m^2<0$ or $0<m^2\leq 2k^2$, then the solution (\ref{a}) dose not meet the first boundary condition. So there is no graviton, {\it i.e.}, $w_{\mu\nu}=u_{\mu\nu}=0$. 

\item $2k^2<m^2\le \frac{9}{4}k^2$, and by using (\ref{c}), $w_{\mu\nu}$~ is proportional to $u_{\mu\nu}$, so to find them, we normalize the solution to one:
\begin{align*}
{}&\int_{-\frac{1}{k}}^{\frac{1}{k}}e^{-\sigma}D_{\mu\nu}^2dy = 1 \nonumber\\
|u_{\mu\nu}|^2\int \Big(F^2(1-k{}&|y|)^{a}  +(1-k|y|)^{-a}-2F\Big)dy=1
\nonumber\\
 |u_{\mu\nu}|^2{}&=\frac{k}{2\Big(\frac{F^2}{1+a}+\frac{1}{1-a}-2F\Big
)}
\end{align*}
Then
\begin{equation}
D_{\mu\nu}^2\Big|_{y=0}=|u_{\mu\nu}|^2(F-1)^2=k\frac{m^2-2k^2}{m^2-k^2}\label{3}
\end{equation}
\item $\frac{9}{4}k^2<m^2$ then $n_{1,2}=(1\pm i|a|)/2$, using  (\ref{c}) and normalizing the solution to one, we find:
\begin{align*}
{}&\int_{-\frac{1}{k}}^{\frac{1}{k}} e^{-\sigma}D_{\mu\nu}^2dy=1
\\
|u_{\mu\nu}|^2\int \Big(|F|^2+1{}&-F(1-k|y|)^{+ia}-F^\ast (1-k|y|)^{-ia}\Big)dy=1\\
|u_{\mu\nu}|^2{}&=\frac{k}{2\Big(|F|^2+1-2Re(\frac{F}{1+ia})\Big
)} \label{4th}
\end{align*}
Then
\begin{equation}
D_{\mu\nu}^2\Big|_{y=0}=|u_{\mu\nu}|^2(|F|^2+1-2Re(F))=k\frac{m^2-2k^2}{m^2-k^2}\label{4}
\end{equation}
\end{enumerate}
Note that the final results in eq.s (\ref{3}) and (\ref{4}) are the same.

Before going on it is worth to comment on these results. Firstly, there is no negative mass-squared graviton. This implies the stability of the background metric against tensorial perturbations. Our analysis reveals that there is a massless mode which as we will see in the next section is important to reach a Newtonian gravity limit. Moreover, there is a mass gap which separates the massless mode from the continuum. This is consistent with the fact that  massive gravitons have not yet been found experimentally.   
\section{Newtonian limit}
The gravitational potential of a source mass $m$ produced by the exchange of graviton reads as Yukawa potential \cite{RS}:
\begin{equation}\label{yukawa}
V(r)=\frac{m}{4\pi}\sum_{m_n=0}^{+\infty}g^2\frac{e^{-m_nr}}{r}
\end{equation}
where  sum is over the graviton spectrum found in the previous section and $g$ is coupling constant for the gravitational interaction and can be given as  \cite{effective},
\begin{equation} 
g^2=\frac{\kappa^2}{4}\tilde{D}^2_{\mu\nu}\Big|_{y=0} = \left\{ \begin{matrix}
\frac{\kappa^2 k}{2} & \;\;\;\;\text{for}\;\;\;\; m^2=0\\
\frac{\kappa^2 k}{4}\left(\frac{m^2-2k^2}{m^2-k^2} \right)   & \;\;\;\;\text{for}\;\;\;\; m^2>2k^2\\
\end{matrix} \right. 
\end{equation}
in which we have used (\ref{1})-(\ref{4}).

The interaction term in gravitational Lagrangian is \cite{effective}:
\begin{equation} 
\mathcal{L}_{int}=\frac{\kappa}{2}h_{\mu\nu}\Big|_{y=0}T^{\mu\nu}
\end{equation} 
where $T^{\mu\nu}$ is the energy-momentum tensor for matters on the 3-brane and $\kappa^2=\frac{8\pi G}{k}$ such that:
\begin{equation} 
S_g=\frac{1}{2\kappa^2}\int \sqrt{|g^{(5)}|}R^{(5)}dydx
\end{equation} 
because $R^{(4)}_{\mu\nu}$~is contained in $R^{(5)}_{\mu\nu}$:
\begin{align*} 
S_g {}&\sim  \frac{1}{2\kappa^2}\int e^{2\sigma}\sqrt{|g^{(4)}|} e^{-\sigma} R^{(4)}dydx +\dots  \\
{}& =\frac{1}{2\kappa^2}\int \sqrt{|g^{(4)}|}R^{(4)}\frac{1}{k}dx
\end{align*}
then:
\begin{equation} 
\frac{1}{2\kappa^2k}=\frac{1}{16\pi G}
\end{equation} 
where $G$ is the 4-dimensional gravitational constant.
Because of partly continuous mass spectrum, sum in (\ref{yukawa}) turns into an integration and it could be revealed that the integral measure $dm$~ is proportionate to $k$ :
\begin{equation}
V(r)=\frac{Gm}{2k}\Big(\frac{2k}{r}+\sum_{m_n\ne 0}^{+\infty}\frac{e^{-m_nr}}{r}\frac{g^2}{k}\Big)=\frac{Gm}{r}\Big(1+\int_{\sqrt2k}^{+\infty}\frac{1}{2k}\frac{m^2-2k^2}{m^2-k^2}e^{-mr}dm \Big) \label{v}
\end{equation}
the integration in right hand side is accounted as:
\begin{align}
\int_{\sqrt2k}^{+\infty}\frac{m^2-2k^2}{m^2-k^2}e^{-mr}dm&=\Big( \frac{2e^{kr}-k Ei\big(1,(\sqrt2-1)kr\big)e^{\sqrt2kr}r+k Ei\big(1,(\sqrt2+1)kr\big)e^{(\sqrt2+2)kr}r}{2e^{(\sqrt2+1)kr}r}\Big) \nonumber\\
&\simeq e^{-2\sqrt2kr}\Big(-\frac{1}{2(\sqrt2-1)r}+O\big(\frac{1}{r^2}\big)\Big) \;\;\;\;\text{for}\;\;\;\; kr\gg 1
\end{align}

Then potential for $r\gg1$~at first orders reads as:
\begin{equation}
V(r)\simeq \frac{Gm}{r}\Big(1-\frac{e^{-2\sqrt2kr}}{4(\sqrt2-1)kr}\Big) \label{p}
\end{equation}
So, as eq. (\ref{p}) justifies, MRS model meets the gravitational Newtonian limit consistent with observations. 

\section{Discussion and conclusion}
In this paper, we constructed the gravitational perturbation of a MRS model (\ref{em}) and found equations of motion for different components of metric perturbation as (\ref{mn}), (\ref{m5}) and (\ref{55}). By the decomposition (\ref{de}) and the unitary gauge, we solved the equations of motion. Mass spectrum for graviton modes depicted a massless mode which is in charge of the Newtonian gravity limit, and a continuum that is responsible for small correction to the Newtonian limit in short distances. Interestingly, there is a mass gap which separates the single massless mode from the mass continuum that could be addressed why massive gravitons have not yet been detected experimentally. The higher dimensional corrections to the Newtonian gravity are exponentially suppressed due to the mass gap. Similar behavior has been recently reported in \cite{German:2015cna}.

Finally it is worth mentioning that there is no negative mass-squared in the graviton spectrum which indicates the stability of the model.  



\end{document}